\documentclass[sigconf]{acmart}

\usepackage{xcolor}
\usepackage{graphicx}
\usepackage{booktabs}
\usepackage{tikz}

\newcommand*\circled[1]{\tikz[baseline=(char.base)]{
            \node[shape=circle,draw,fill=black,inner sep=1pt,font=\small] (char) {\textcolor{white}{#1}};}}

\AtBeginDocument{%
  }

\setcopyright{acmlicensed}
\copyrightyear{2026}
\acmYear{2026}
\acmDOI{XXXXXXX.XXXXXXX}
\acmConference[ASE '26]{41st IEEE/ACM International Conference on Automated Software Engineering}{October 12--16, 2026}{Munich, Germany}
\acmBooktitle{41st IEEE/ACM International Conference on Automated Software Engineering (ASE '26), October 12--16, 2026, Munich, Germany}
\acmISBN{978-1-4503-XXXX-X/2018/06}

\begin{document}

\emergencystretch=3em

\title{FeatX: Editing Software by Editing Features for Repository-Level Code Evolution}

\author{Xutian Li}
\email{xtli25@stu.pku.edu.cn}
\orcid{0009-0001-2441-7447}
\affiliation{%
  \institution{Peking University}
  \city{Beijing}
  \country{China}
}

\author{Yifeng Zhu}
\email{yifengzhu25@stu.pku.edu.cn}
\orcid{0009-0009-0487-8404}
\affiliation{%
  \institution{Peking University}
  \city{Beijing}
  \country{China}
}

\author{Xianlin Zhao}
\email{zhaoxianlin@pku.edu.cn}
\orcid{0009-0007-1471-7685}
\affiliation{%
  \institution{Peking University}
  \city{Beijing}
  \country{China}
}

\author{Yanzhen Zou}
\email{zouyz@pku.edu.cn}
\orcid{0009-0009-1764-6645}
\affiliation{%
  \institution{Peking University}
  \city{Beijing}
  \country{China}
}

\author{Lu Zhang}
\email{zhanglucs@pku.edu.cn}
\orcid{0000-0001-8304-7055}
\affiliation{%
  \institution{Peking University}
  \city{Beijing}
  \country{China}
}

\author{Bing Xie}
\email{xiebing@pku.edu.cn}
\orcid{0000-0002-2988-2575}
\affiliation{%
  \institution{Peking University}
  \city{Beijing}
  \country{China}
}
\renewcommand{\shortauthors}{Li et al.}

\begin{abstract}
Large language models (LLMs) are increasingly used for software evolution, yet most interaction paradigms remain code-centric and require manual context management and prompt iteration.
We present FeatX, a feature-oriented tool for editing software by editing features.
Given an existing repository, FeatX extracts a hierarchical epic-feature structure with explicit feature-to-code mappings, then invokes a three-stage Evolution Agent to translate feature edits into code patches.
The workflow is exposed through four coordinated panels.
Across a controlled user study and replay experiments on 38 real-world feature-editing commits, FeatX significantly reduces cognitive load and improves usability compared with vanilla ChatGPT. It also achieves a 42.6\% relative improvement in function-level modification localization F1 over strong LLM baselines, at substantially lower cost (\$0.07 in total).
The tool and collected dataset are available at \url{https://github.com/a496263365/FeatX/tree/demo}, with a demonstration video at \url{https://youtu.be/OZqKZ4Ii-yM}. 

\end{abstract}

\begin{CCSXML}
<ccs2012>
    <concept>
        <concept_id>10011007.10011074.10011111.10011696</concept_id>
        <concept_desc>Software and its engineering~Maintaining software</concept_desc>
        <concept_significance>500</concept_significance>
    </concept>
    <concept>
        <concept_id>10011007.10011074.10011111.10011113</concept_id>
        <concept_desc>Software and its engineering~Software evolution</concept_desc>
        <concept_significance>500</concept_significance>
    </concept>
    <concept>
        <concept_id>10011007.10011074.10011092.10011782</concept_id>
        <concept_desc>Software and its engineering~Automatic programming</concept_desc>
        <concept_significance>500</concept_significance>
    </concept>
    <concept>
        <concept_id>10003120.10003121.10003129</concept_id>
        <concept_desc>Human-centered computing~Interactive systems and tools</concept_desc>
        <concept_significance>500</concept_significance>
    </concept>
 </ccs2012>
\end{CCSXML}

\ccsdesc[500]{Software and its engineering~Maintaining software}
\ccsdesc[500]{Software and its engineering~Software evolution}
\ccsdesc[500]{Software and its engineering~Automatic programming}
\ccsdesc[500]{Human-centered computing~Interactive systems and tools}

\keywords{Feature-Level Abstraction, Software Evolution, Large Language Models, Developer Interface}

\maketitle

\section{Introduction}

Software evolution accounts for a substantial portion of software engineering effort.
Unlike single-function generation tasks, practical evolution requires translating repository-level \textit{feature intent} into coordinated \textit{code edits} across multiple files.
Prior studies~\cite{922739} indicate that around 60\% of repository maintenance tasks involve feature evolution.

Large language models (LLMs) are increasingly used for software evolution, and existing approaches mainly provide four interaction paradigms.
Autocompletion tools (e.g., GitHub Copilot inline suggestions) improve local editing efficiency.
Chat-based workflows (e.g., Cursor, SWE-Agent~\cite{swe-agent}) support natural-language intent expression for evolution tasks.
Annotation-based approaches (e.g., NL Outlines~\cite{outlines}, Code Shaping~\cite{codeshaping}) strengthen local intent-code alignment.
Design-oriented systems (e.g., \textsc{Pail}~\cite{pail}, EvoDev~\cite{EvoDev}) improve planning quality for development from scratch.

However, these paradigms still provide limited support for repository-level feature evolution in existing repositories.
They typically do not provide developers with an explicit feature list, making task descriptions ambiguous and inconsistent with the its actual structure.
They also do not maintain a feature-to-code mapping, which reduces localization accuracy during requirement analysis and context retrieval.
This gap is especially costly in unfamiliar repositories: developers must infer features and manually map intended feature changes to concrete edit locations, which imposes substantial cognitive overhead~\cite{miller1956}.
For example, a commit adding \textit{friend links} in NBlog\footnote{\url{https://github.com/Naccl/NBlog/commit/d47882e69d318b82827ada83b197742bec4d9669}} spans 5 packages, 5 files, and 28 functions.

To address these problems, we present \textbf{FeatX}, a feature-oriented software evolution tool that supports editing software by editing features.
FeatX integrates automatic feature extraction with an Evolution Agent. It treats extracted features as first-class editable units and maintains explicit feature-to-code mappings throughout generation.
Given an existing repository, FeatX first constructs a hierarchical epic-feature structure with associated code entities, and then invokes a three-stage Evolution Agent (contextual expansion, localization \& planning, and concrete code modification) to translate feature edits into code patches.
The workflow is exposed through four coordinated panels (Feature Panel, CodeMap Panel, Agent Panel, and Diff Panel).

\begin{figure*}[!t]
    \centering
    \includegraphics[width=0.8\textwidth]{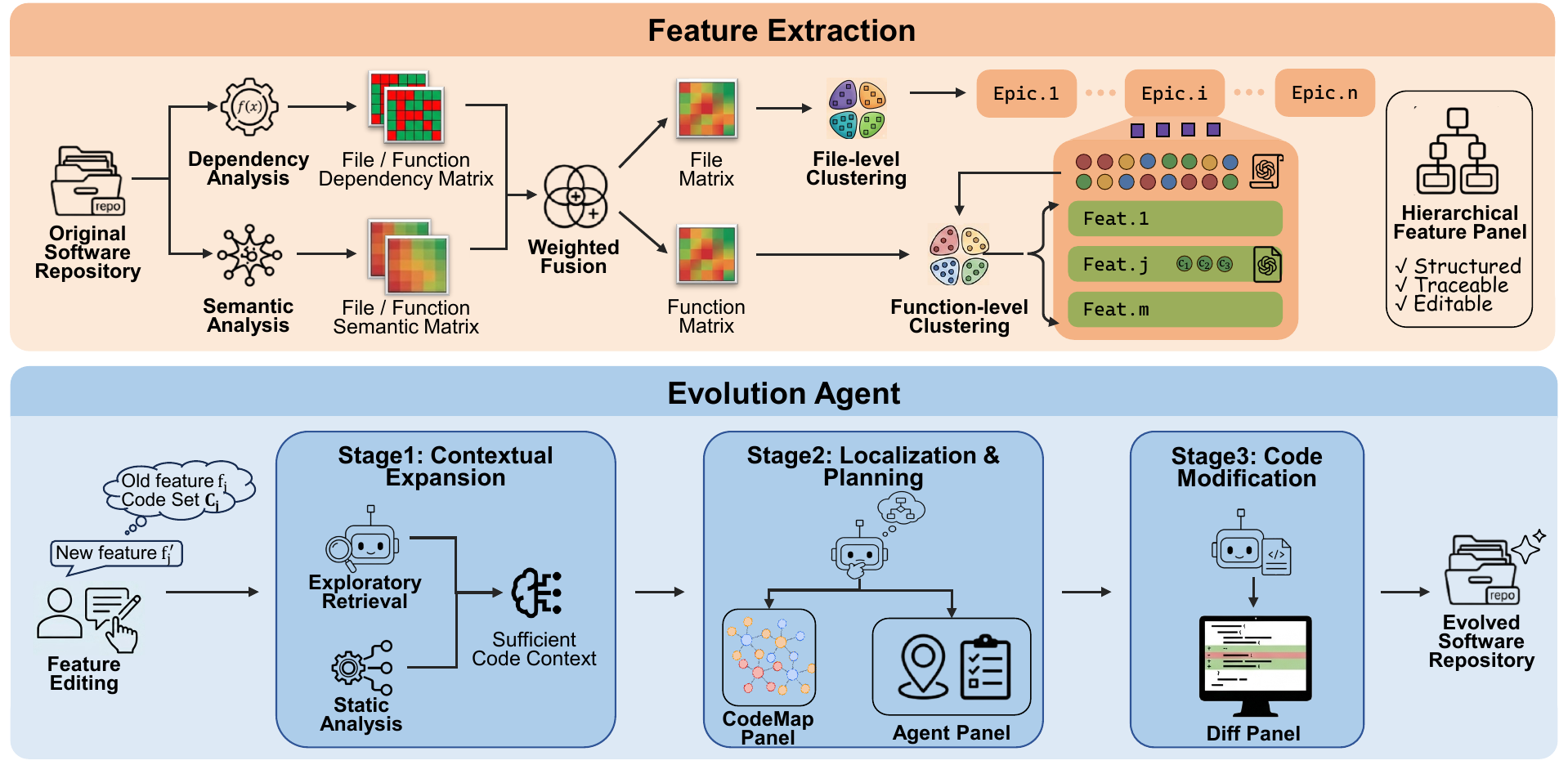}
    \caption{The overall design of FeatX.}
    \label{fig:overview}
\end{figure*}

We evaluate FeatX along two dimensions: cognitive load and function-level modification localization accuracy.
Across a user study and quantitative experiments on real-world feature-editing tasks, FeatX shows that: 
(1) compared with vanilla ChatGPT, it significantly reduces NASA-TLX cognitive load ($12.5 \to 7.4$, 41\% reduction) and improves usability ($73 \to 84$, 15\% increase); and
(2) it improves function-level modification localization accuracy, achieving an F1 score of 0.385 and outperforming the strongest baseline, Claude-opus-4.5 (0.270), by 42.6\%. Meanwhile, it only incurs an LLM cost of \$0.07, in contrast to \$45.05 for direct Claude invocation.

\section{Tool Design}

\autoref{fig:overview} presents the overall design of FeatX. 
It extracts a hierarchical feature structure from a repository (\autoref{sec:extraction}), lets users edit target features in the Feature Panel, and feeds the edited feature with associated code context to a three-stage Evolution Agent that outputs diffs (\autoref{sec:agent}).

\subsection{Feature Extraction}
\label{sec:extraction}

\subsubsection{Problem Formulation}
We adopt a hierarchical requirement model inspired by Agile/Scrum~\cite{agile}: each \textit{Epic} contains multiple \textit{Features}, and each feature is represented by a user story and supporting code entities.

Given a repository $R$, we construct a two-level hierarchy $E_R=\{e_1,\dots,e_n\}$, where each epic $e_i$ contains features $F_i=\{f_1,\dots,f_m\}$ and each feature $f_j$ is associated with a user story $s_j$ and code entities $C_j$ (classes, methods, or files).

\subsubsection{Extraction Pipeline}
In our previous work~\cite{reposummary}, we introduced a feature-oriented code repository summarization approach.
Here, we adapt this approach for feature extraction to build the hierarchical feature list used in FeatX.
For both file-level and function-level entities, we first construct a binary structural matrix from static dependencies (e.g., imports and calls), and a continuous semantic matrix in $[0,1]$ from LLM-based summaries and Sentence-BERT embeddings.
We then fuse the two matrices with weighted integration to obtain a unified similarity matrix at each granularity.

Next, we perform two-stage hierarchical clustering using the Leiden algorithm.
Within the whole repository, we cluster the fused file-level matrix and ask the LLM to generate one description for each cluster as an \textit{epic}.
Within each epic, we further cluster the corresponding fused function-level representations and ask the LLM to generate descriptions for the resulting clusters as \textit{features}.
The final epic-feature hierarchy is shown in the Feature Panel for user editing.

\begin{figure*}[!t]
    \centering
    \includegraphics[width=0.8\textwidth]{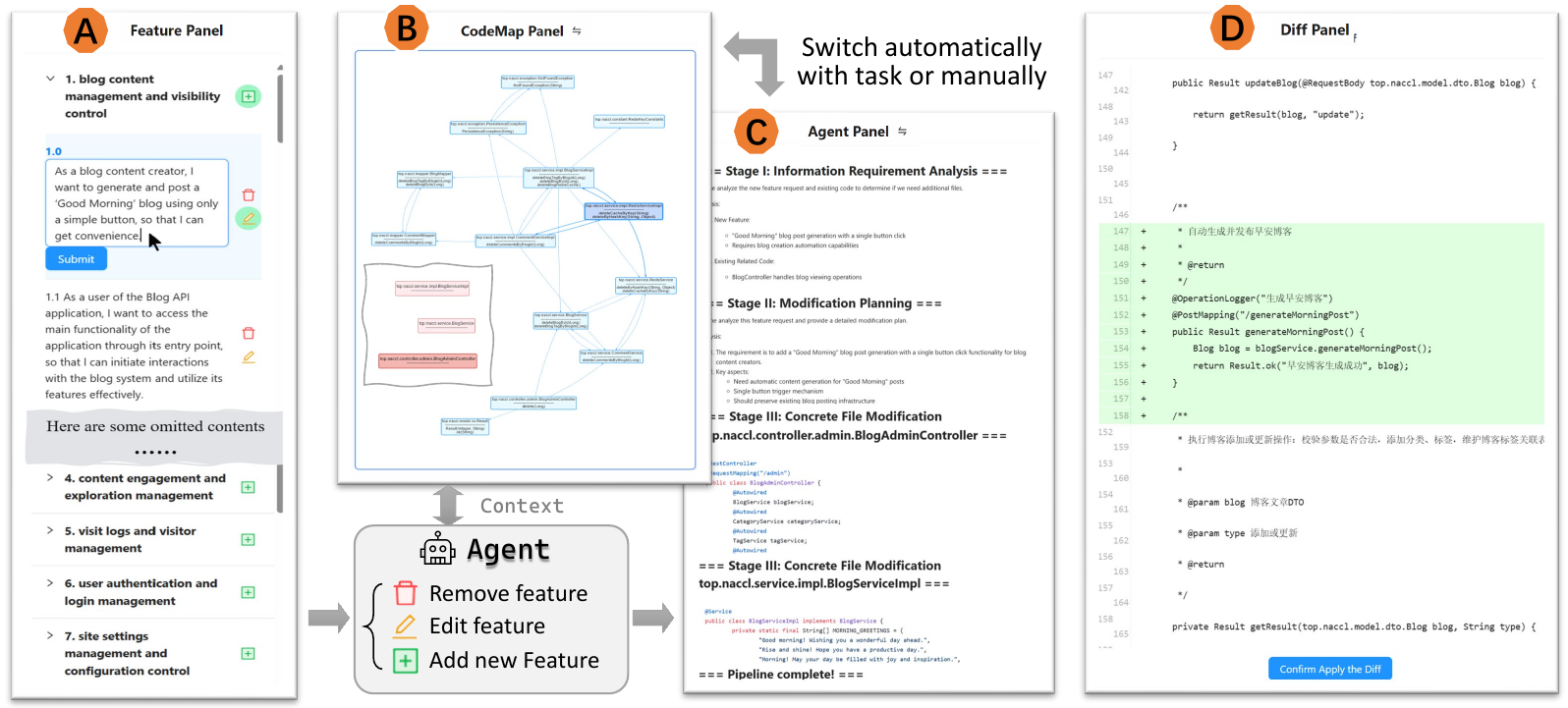}
    \caption{Usage workflow of FeatX.}
    \label{fig:usage}
\end{figure*}

\subsection{Evolution Agent}
\label{sec:agent}

FeatX implements a three-stage Evolution Agent to transform feature-level edits into code-level modifications.
Given a user-edited feature, the agent also receives the original feature description and its associated code context, then executes contextual expansion, localization \& planning, and code modification in sequence~\cite{agentless}.

\subsubsection{Contextual Expansion}

To provide sufficient implementation context, this stage expands the initial code entities linked to the edited feature from two complementary perspectives: \textit{Static Analysis} and \textit{Exploratory Retrieval}.

In the \textit{Static Analysis} branch, FeatX starts from the seed entities identified in \autoref{sec:extraction} and expands them along repository-level dependency relations~\cite{repograph}, including \textit{method calls}, \textit{field use references}, \textit{field definitions}, \textit{type use references}, \textit{class membership}, and \textit{inheritance/nesting hierarchies}.
In the \textit{Exploratory Retrieval} branch, FeatX leverages LLM-guided exploration to introduce additional semantically relevant entities that may not be reachable through static dependency expansion.

The result is an enriched CodeMap for downstream reasoning, visualized in the CodeMap Panel.

\subsubsection{Localization and Planning}
This stage compares the feature description before and after user edits, and combines the semantic delta with the expanded code context.
Based on this joint signal, the agent localizes the change intent to a small set of concrete code regions.
It then generates a fine-grained modification plan for each localized region, including required edits, dependencies, and execution order.
The reasoning process and planning results are displayed in the Agent Panel.

\subsubsection{Concrete Code Modification}
This stage modifies concrete code entities according to localized targets and plans.
The output is produced as class-wise line-level diffs in the Diff Panel for review and confirmation.
In addition, affected classes are highlighted in the CodeMap Panel.

This three-stage design makes feature editing operational in practice: user intent is first grounded into an enriched CodeMap, then translated into explicit edit plans, and finally materialized as reviewable diffs with affected entities highlighted in the interface.

\section{Implementation and Usage}

FeatX is implemented as a decoupled web system with a Java Spring Boot backend and a JavaScript React frontend.
To support software evolution in Java repositories, we use JavaParser~\cite{JavaParser} for static analysis and dependency resolution.
In feature extraction, we set the fusion weight between the structural and semantic matrices to $\alpha=0.5$, and apply adaptive tuning to select the best $\gamma$ for cluster partitioning modularity (CPM).
Across the system, all LLM-dependent components, including feature extraction and Evolution Agent, use the same open-source \texttt{DeepSeek-v3} configuration with temperature 0.0 to ensure deterministic behavior.

As shown in \autoref{fig:usage}, FeatX provides a concise feature-oriented workflow through four coordinated panels:
\circled{A} the Feature Panel lists extracted epics and features hierarchically and supports direct addition, modification, and removal operations;
\circled{B} the CodeMap Panel visualizes feature-related code entities and highlights entities affected by the current edit;
\circled{C} the Agent Panel presents the three-stage reasoning process of Evolution Agent;
and \circled{D} the Diff Panel provides class-wise line-level code diffs for final review and confirmation.
This feature-oriented workflow shifts user effort from manual repository navigation and context management to feature specification and targeted verification, improving traceability between feature intent and concrete code evolution.

\section{Evaluation}

To evaluate FeatX, we investigate the following research questions:

\begin{itemize}
    \item[\textbf{RQ1.}] From a user perspective, does FeatX reduce cognitive load and improve usability compared with vanilla ChatGPT?
    \item[\textbf{RQ2.}] What function-level modification localization accuracy and cost does FeatX achieve?
\end{itemize}

The evaluation uses 38 feature-editing commits from five open-source Java repositories (\textit{FlappyBird}, \textit{JSON-java}, \textit{JCommander}, \textit{NBlog}, and \textit{PlayEdu})~\cite{reposummary}, with average complexity of 3.2 packages, 4.5 files, and 10.6 functions per task.

\subsection{User-Perceived Interaction Quality}

We recruited 10 participants (6 men and 4 women; ages 21--27), including 8 CS students (undergraduate to PhD) from 3 universities and 2 industry practitioners.
Participants were recruited via snowball sampling and generally had prior Java/OOP experience and regular LLM usage experience.
Each participant completed two medium-complexity feature-editing tasks selected from our commit dataset using both vanilla ChatGPT and FeatX. 
Each task involved modifications to 5--10 functions across multiple files, and each session was limited to one hour.
We measured NASA-TLX~\cite{nasa-tlx} and SUS~\cite{sus} with two-sided paired t-tests, and report only mean scores and significance levels due to page limits.

\subsubsection{Cognitive Load Reduction}
\begin{figure}[htbp]
    \centering
    \includegraphics[width=0.95\linewidth,trim=10 10 0 0,clip]{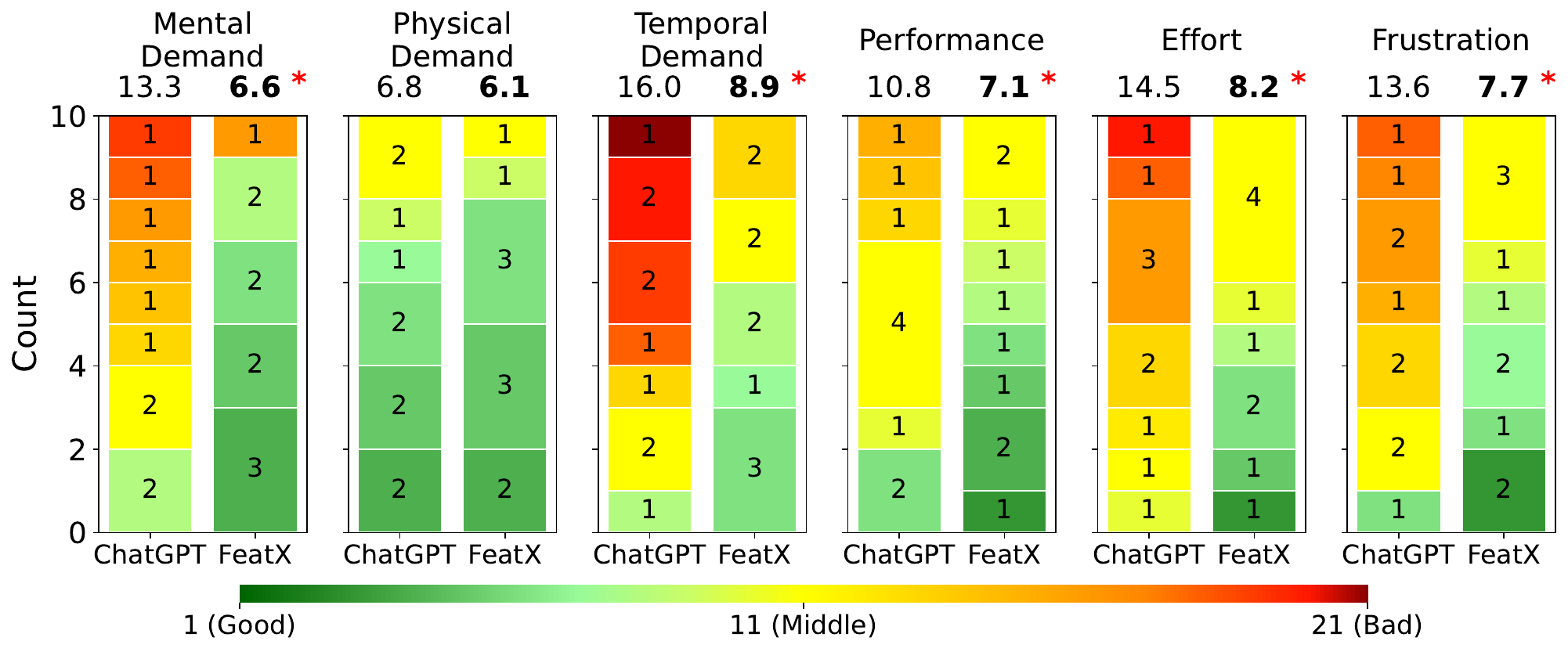}
    \caption{NASA-TLX cognitive load ratings.}
    \label{fig:us}
\end{figure}

NASA-TLX rates workload on six dimensions using a 1--21 scale, where higher scores indicate higher cognitive load.
As shown in \autoref{fig:us}, FeatX significantly reduces perceived cognitive load compared with ChatGPT, 
with the largest reductions in \textit{mental demand} (6.6 vs.\ 13.3), \textit{effort} (8.2 vs.\ 14.5), and \textit{frustration} (7.7 vs.\ 13.6) (all $p < 0.001$).
The only dimension without a significant difference is \textit{physical demand} ($p > 0.05$).
Averaged across all six dimensions, the mean NASA-TLX score decreases from 12.5 to 7.4 (41\% reduction).

\subsubsection{Usability Improvement}

\begin{figure}[htbp]
    \centering
    \includegraphics[width=0.8\linewidth,trim=16 16 0 0,clip]{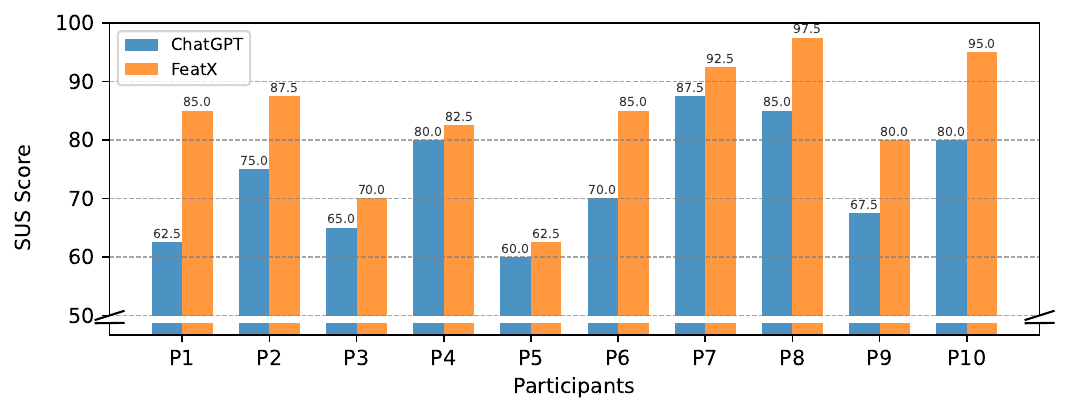}
    \caption{Individual SUS scores comparison.}
    \label{fig:sus}
\end{figure}

SUS consists of 10 items assessing overall system usability.
\autoref{fig:sus} reports each participant's final SUS score, showing a higher mean score for FeatX than ChatGPT (84 vs.\ 73, 15\% increase).
At the item level, two-sided paired t-tests on the 10 SUS questions show that FeatX does not outperform ChatGPT on every item, but scores significantly higher on \textit{integration} and \textit{consistency} ($p < 0.05$), and especially on \textit{confidence} ($p < 0.001$).

\subsection{Function-Level Modification Localization Accuracy and Cost}

We evaluate localization at the function level. 
For each task, we compare a ground-truth modified-function set $G$ from commit history with a predicted set $M$.
We compute $|M\cap G|$ as follows: function names are matched by exact match (EM), while newly created functions are manually assessed by the authors for correctness. 
We then compute $\text{Precision}=\frac{|M\cap G|}{|M|}$, $\text{Recall}=\frac{|M\cap G|}{|G|}$, and $\text{F1}=\frac{2\cdot \text{Precision}\cdot \text{Recall}}{\text{Precision}+\text{Recall}}$.

\begin{table}
\caption{Results of localization accuracy and cost.}
\label{tab:quantitative}
\centering
\resizebox{0.9\columnwidth}{!}{%
\small
\begin{tabular}{lcccc}
\toprule
\textbf{Approaches} & \textbf{Precision (\%)} & \textbf{Recall (\%)} & \textbf{F1 Score} & \textbf{Cost (\$)} \\
\midrule
DeepSeek-v3 & 36.0 & 9.0 & 0.143 & - \\
DeepSeek-v3.2-think & 29.7 & 5.5 & 0.092 & - \\
GPT-4o-mini & 37.0 & 9.7 & 0.154 & 1.28 \\
GPT-5.2 & 30.0 & 6.0 & 0.100 & 17.81 \\
Claude-opus-4.5 & \textbf{50.7} & \underline{18.4} & \underline{0.270} & 45.05 \\
Cursor (Agent) & 15.9 & 13.0 & 0.143 &  - \\
\midrule
FeatX (Ours) & \underline{41.6} & \textbf{35.8} & \textbf{0.385} & \textbf{0.07} \\
\bottomrule
\end{tabular}
}
\end{table}

We replayed 38 commits with FeatX, several SOTA LLMs, and Cursor (Agent mode, default model \texttt{composer-1.5}), and reported function-level localization metrics and cost in \autoref{tab:quantitative}.
The prompting protocol is accessible in our repository.
FeatX achieved the highest F1 (0.385), with 41.6\% precision and 35.8\% recall, while the second-best F1 was achieved by Claude-opus-4.5 (0.270), corresponding to a 42.6\% relative improvement.
Despite being widely used in practice, Cursor Agent attains only 13.0\% recall and has the lowest precision (15.9\%).
In addition, FeatX incurs lower LLM cost (\$0.07 in total) than direct Claude invocation (\$45.05).

\section{Conclusion and Future Work}
\label{sec:conclusion}

We present FeatX, a feature-oriented tool for LLM-assisted software evolution.
FeatX enables editing software by editing features, with explicit feature extraction and a three-stage Evolution Agent.
Our evaluation shows improvements in user-perceived interaction quality and function-level modification localization, together with strong cost efficiency.
In future work, we will improve robustness in complex edits, support interactive relevant-code augmentation, and evaluating on larger repositories with more diverse participants.

\bibliographystyle{ACM-Reference-Format}
\bibliography{reference}

\end{document}